\documentclass[
reprint,
superscriptaddress,
amsmath,amssymb,
apl,
]{revtex4-2}

\usepackage[utf8]{inputenc}
\usepackage{graphicx,import}
\graphicspath{{Figures/}}
\usepackage{bm}
\usepackage[colorlinks]{hyperref}
\usepackage[all]{hypcap}
\usepackage{xcolor}
\usepackage{braket}
\usepackage{siunitx}
\usepackage{changes}
\usepackage{lipsum}

\definecolor{bluegray}{RGB}{40,180,160}
\definecolor{navygray}{RGB}{110,140,170}
\definecolor{meadowgreen}{RGB}{0,128,0}
\definecolor{coolbrown}{RGB} {165,42,42}

\DeclareSIUnit{\sq}{\Box}

\newcommand{\secref}[1]{\hyperref[#1]{{Section~\ref{#1}}}}
\newcommand{\chapref}[1]{\hyperref[#1]{{Chapter~\ref{#1}}}}
\newcommand{\suppref}[1]{\hyperref[#1]{{App.~\ref{#1}}}}
\newcommand{\figref}[1]{\hyperref[#1]{{Fig.~\ref*{#1}}}}
\newcommand{\Figref}[1]{\hyperref[#1]{{Figure~\ref*{#1}}}}
\newcommand{\figrefadd}[2]{\hyperref[#1]{{Fig.~\ref*{#1}#2}}}
\newcommand{\Figrefadd}[2]{\hyperref[#1]{{Figure~\ref*{#1}#2}}}
\newcommand{\tabref}[1]{\hyperref[#1]{\textsl{Table~\ref*{#1}}}}
\renewcommand{\eqref}[1]{\hyperref[#1]{{Eq.~\ref*{#1}}}}

\newcommand{\revise}[1]{{#1}}
\newcommand{\revisex}[1]{{}}
\newcommand{\unrevise}[1]{{#1}}

\hypersetup{
citecolor={navygray}, 
linkcolor={navygray},
urlcolor={navygray},
}

\makeatletter
\newcommand*{\balancecolsandclearpage}{%
  \close@column@grid
  \cleardoublepage
  \twocolumngrid
}
\makeatother

\begin{document}
\title{Gralmonium: Granular Aluminum Nano-Junction Fluxonium Qubit}

\author{D.~Rieger}
\email{dennis.rieger@kit.edu}
\affiliation{PHI,~Karlsruhe~Institute~of~Technology,~76131~Karlsruhe,~Germany}

\author{S.~Günzler}
\thanks{First two authors contributed equally.}
\affiliation{PHI,~Karlsruhe~Institute~of~Technology,~76131~Karlsruhe,~Germany}
\affiliation{IQMT,~Karlsruhe~Institute~of~Technology,~76344~Eggenstein-Leopoldshafen,~Germany} 

\author{M.~Spiecker}
\affiliation{PHI,~Karlsruhe~Institute~of~Technology,~76131~Karlsruhe,~Germany}

\author{P.~Paluch}
\affiliation{PHI,~Karlsruhe~Institute~of~Technology,~76131~Karlsruhe,~Germany}
\affiliation{IQMT,~Karlsruhe~Institute~of~Technology,~76344~Eggenstein-Leopoldshafen,~Germany}

\author{P.~Winkel}
\affiliation{PHI,~Karlsruhe~Institute~of~Technology,~76131~Karlsruhe,~Germany}
\affiliation{IQMT,~Karlsruhe~Institute~of~Technology,~76344~Eggenstein-Leopoldshafen,~Germany} 

\author{L.~Hahn}
\affiliation{IMT,~Karlsruhe~Institute~of~Technology,~76344~Eggenstein-Leopoldshafen,~Germany}

\author{J.~K.~Hohmann}
\affiliation{IMT,~Karlsruhe~Institute~of~Technology,~76344~Eggenstein-Leopoldshafen,~Germany}

\author{A.~Bacher}
\affiliation{IMT,~Karlsruhe~Institute~of~Technology,~76344~Eggenstein-Leopoldshafen,~Germany}

\author{W.~Wernsdorfer}
\affiliation{PHI,~Karlsruhe~Institute~of~Technology,~76131~Karlsruhe,~Germany}
\affiliation{IQMT,~Karlsruhe~Institute~of~Technology,~76344~Eggenstein-Leopoldshafen,~Germany}

\author{I.~M.~Pop}
\email{ioan.pop@kit.edu}
\affiliation{PHI,~Karlsruhe~Institute~of~Technology,~76131~Karlsruhe,~Germany}
\affiliation{IQMT,~Karlsruhe~Institute~of~Technology,~76344~Eggenstein-Leopoldshafen,~Germany}

\date{\today}

\begin{abstract}
Mesoscopic Josephson junctions (JJs), consisting of overlapping superconducting electrodes separated by a nanometer thin oxide layer, provide a precious source of nonlinearity for superconducting quantum circuits and are at the heart of state-of-the-art qubits, such as the transmon and fluxonium. Here, we show that in a fluxonium qubit the role of the JJ can also be played by a lithographically defined, self-structured granular aluminum (grAl) nano-junction: a superconductor-insulator-superconductor (SIS) JJ obtained in a single layer, zero-angle evaporation. The measured spectrum of the resulting qubit, which we nickname gralmonium, is indistinguishable from the one of a standard fluxonium qubit. Remarkably, the lack of a mesoscopic parallel plate capacitor gives rise to an intrinsically large grAl nano-junction charging energy in the range of \revise{tens of \si{\giga\hertz}}, comparable to its Josephson energy  $E_\text{J}$. We measure average energy relaxation times of $T_1=\SI{10}{\micro\second}$ and Hahn echo coherence times of $T_2^\text{echo}=\SI{9}{\micro\second}$. The exponential sensitivity of the gralmonium to the $E_\text{J}$ of the grAl nano-junction provides a highly susceptible detector. Indeed, we observe spontaneous jumps of the value of $E_\text{J}$ on timescales from milliseconds to days, which offer a powerful diagnostics tool for microscopic defects in superconducting materials.
\end{abstract}

\maketitle
The remarkable progress of superconducting quantum information processing in academia~\cite{Nakamura1999, Vion2002, 2003FluxQ, transmon, Lupascu2007, fluxonium, Gu2017QubitReview,  Carusotto2020,  Blais2021CQED} and in\revisex{ companies} \revise{industry}~\cite{Kandala2019, Gold2021Sep,alibaba2021fluxonium, McEwen2022} is fueled by the Josephson effect~\cite{josephson1962}, which provides nonlinearity while maintaining coherence. In practice, the vast majority of JJs are implemented in the form of Al/AlO$_x$/Al SIS weak links, which offer a long list of benefits, such as high control over the insulating barrier~\cite{Kreikebaum2020Apr}, robustness to thermal cycling~\cite{3Dtransmon2011} and unmatched coherence~\cite{3Dtransmon2011, tantalumTransmon, somoroff2021millisecond, Siddiqi2021}. However, due to the involved multi-layer and often multi-angle evaporation processes, it is difficult to reduce the footprint of such mesoscopic SIS JJs significantly below $100\times\SI{100}{\square\nano\meter}$. As a consequence, their critical current is suppressed in a Fraunhofer pattern for magnetic fields in the $10^2\,\si{\milli\tesla}$ range~\cite{andreTransmon,teslatransmon}, which are interesting for hybrid architectures~\cite{Bonizzoni2017, Clerk2020Mar, Kurizki3866}. In addition, the parallel plate electrodes of the JJ entail an unavoidable capacitance, which, if removed, could lift one of the constraints in engineering Hamiltonians with large quantum phase fluctuations~\cite{fluxonium, Bell2014Apr, Groszkowski2018Apr, zeropihouck, Peruzzo2021Nov}.

Other types of weak links can be used to overcome some of these limitations. For example, devices using super-\slash semi-\slash superconductor junctions~\cite{Janvier2015Sep, Hays2021Jul} provide a promising quantum information platform and have been demonstrated to be resilient to magnetic field~\cite{gatemon1, gatemon2, gatefluxonium}, but they introduce additional complexity in the fabrication process and additional noise sources due to the required gate bias. Alternatively, defining a constriction in a superconducting wire to form an ScS JJ harvests nonlinearity directly from a continuous superconducting film~\cite{tinkham, JJreview2004}. While ScS JJs are well established in direct current devices operated in magnetic field~\cite{nanoSQUID, WolfgangScS}, embedding them in qubits made of homogeneously disordered superconductors\revise{~\cite{mooij2005phase, astafiev2012, astafiev2013, astafiev2016, deGraaf2018Jun}} currently yields orders of magnitude higher dissipation compared to SIS JJ qubits~\cite{alibaba2021fluxonium, somoroff2021millisecond, gralfluxonium, schusterLowFreqFluxonium, Groszkowski2018Apr, zeropihouck}.

In this article, we combine the advantageous coherence of SIS JJ with the nanoscopic, single-layer design of ScS JJ by utilizing the self-structured aluminum grain assembly of granular aluminum (grAl)~\cite{grainsize, GlezerMoshe2020Aug} to form a nano-junction. To assess the nonlinearity and coherence of the grAl nano-junction, we incorporate it into a fluxonium superconducting qubit, which we nickname gralmonium. Notably, this allows for a single layer fabrication of the whole circuit, since all circuit elements for qubit and readout can be engineered by only tailoring the geometry of the wires. We demonstrate that the gralmonium follows the standard fluxonium Hamiltonian and measure energy relaxation and coherence times on the order of \SI{10}{\micro\second}, comparable to \revise{many} superconducting qubits based on conventional SIS JJs. Moreover, the magnetic field resilience of grAl~\cite{kiril} and the nanoscopic footprint of the grAl nano-junction make the gralmonium an attractive platform for prospective applications in hybrid quantum architectures. The high susceptibility of the nano-junction to the grAl microstructure is passed on to the gralmonium and could be harnessed in future detector circuits. \revisex{The nano-junction concept presented here has the potential to replace SIS junctions in many devices such as parametric amplifiers, detectors and qubits.}

\begin{figure}[t!]
\centering
\includegraphics[width=\columnwidth]{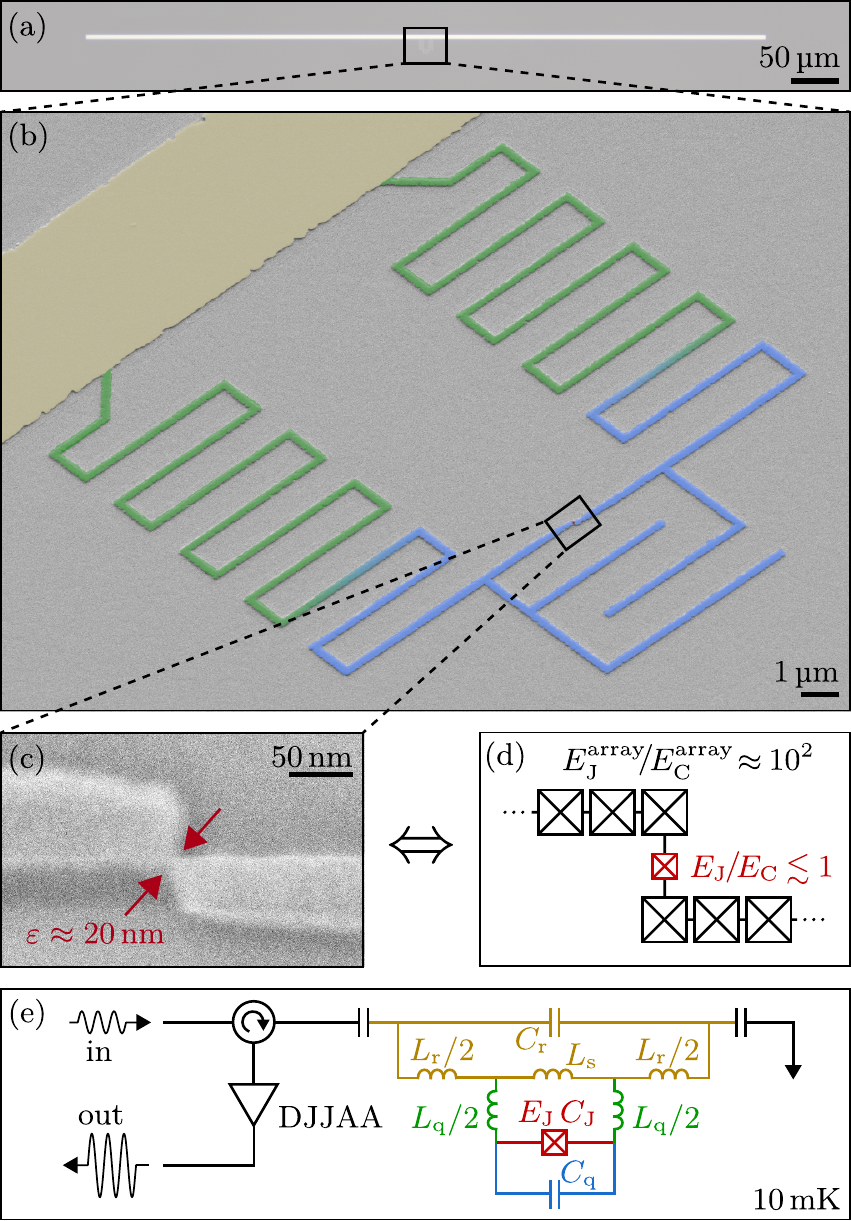}
\caption{\textbf{The gralmonium: a single layer granular aluminum (grAl) fluxonium circuit.} \textbf{(a)} Optical microscope image of the antenna, serving as readout resonator for the gralmonium qubit located in the center. \textbf{(b)} False-colored scanning electron microscope (SEM) zoom-in on the qubit circuit, which consists of a meandered grAl superinductor loop~\cite{gralfluxonium} (green), coupled galvanically to the antenna (ocher) and closed by a coplanar capacitor (blue) in parallel with the grAl nano-junction. \textbf{(c)} Zoom-in on the grAl nano-junction (red arrows) implemented by an $\varepsilon^3$ grAl volume, where $\varepsilon\approx\SI{20}{\nano\meter}$. \textbf{(d)} Effective 1D JJ array circuit model of the grAl nano-junction connected to the superinductor.\label{fig:junction_element} \textbf{(e)} Lumped element circuit schematic of the gralmonium qubit inductively coupled to the readout antenna. The colors of the nano-junction, shunt capacitor $C_\text{q}$, superinductor $L_\text{q}$, resonator capacitor $C_\text{r}$, inductor $L_\text{r}$ and shared inductor $L_\text{s}$ correspond to the highlight colors in panel (b). The readout resonator is measured in single-port reflection using a Dimer Josephson Junction Array Amplifier (DJJAA)~\cite{DJJAA}.
}
\label{fig:sample}
\end{figure}

\Figref{fig:sample} shows the grAl nano-junction embedded into the gralmonium circuit, and the corresponding circuit model. We pattern the readout resonator stripline antenna (\figrefadd{fig:sample}{a}) and the gralmonium located in its center (\figrefadd{fig:sample}{b}) from a \SI{20}{\nano\meter} thick grAl film with sheet resistance \SI{1.5}{\kilo\ohm/\sq} on a sapphire substrate (cf. \suppref{sec:supp:fabrication}). The \SI{170}{\nano\meter} wide meandered superinductance wire shares \SI{8}{\micro\meter} with the antenna, coupling the qubit to the readout similarly to Ref.~\cite{Kou2018Jun}. The flux loop is closed by constricting the wire to an $\varepsilon^3$ grAl volume, with $\varepsilon = \SI{20}{\nano\meter}$: the grAl nano-junction (\figrefadd{fig:sample}{c}). Considering the $\approx\SI{4}{\nano\metre}$ size of the grains in grAl~\cite{grainsize} and the coherence length of our grAl film $\SI{5}{\nano\meter}<\xi<\SI{10}{\nano\meter}<\varepsilon$~\cite{gralCoherence1,Voss2021}, the nano-junction is composed of a 3D network of JJs. However, for frequencies well below the plasma frequency ($\approx \SI{70}{\giga\hertz}$~\cite{natasha, gralCollModes}), the nano-junction and the connecting wire can be modeled as an effective 1D array of SIS JJs~\cite{natasha} with abruptly modulated Josephson coupling (\figrefadd{fig:sample}{d}). Although it is possible that several successive AlO$_x$/Al interfaces contribute to its Josephson coupling, we use a single, effective Josephson energy $E_\text{J}$ and capacitance $C_\text{J}$ to model the nano-junction as a zero dimensional SIS JJ with sinosoidal current phase relation~\cite{JJreview2004}.

In contrast to the superinductor JJs in which $E_\text{J}^\text{array}/ E_\text{C}^\text{array}\approx 10^2$, the nano-junction operates in the opposite regime of $E_\text{J} / E_\text{C} \lesssim 1$, due to its decreased Josephson coupling and small intrinsic capacitance $C_\text{J} < \SI{1}{\femto\farad}$. In order to engineer the total charging energy of the gralmonium to $E_\text{C}^{\Sigma} \approx E_\text{J}$, we add a coplanar capacitor $C_\text{q}$ (highlighted in blue in \figrefadd{fig:sample}{b,e}) in parallel to the nano-junction. Due to the compact geometry of the gralmonium loop, some of the meanders in the vicinity of the junction also contribute to $C_\text{q}$, as illustrated in \figrefadd{fig:sample}{b} by the color gradient green $\to$ blue. We model the entire circuit using an effective lumped element representation depicted in \figrefadd{fig:sample}{e}. The sample is mounted in a sub-wavelength copper tube (see \suppref{sec:supp:sampleholder} and Ref.~\cite{kiril}) and measured in single port microwave reflection employing a parametric quantum amplifier~\cite{DJJAA}.

\begin{figure*}[t!]
\centering
\includegraphics[width=\textwidth]{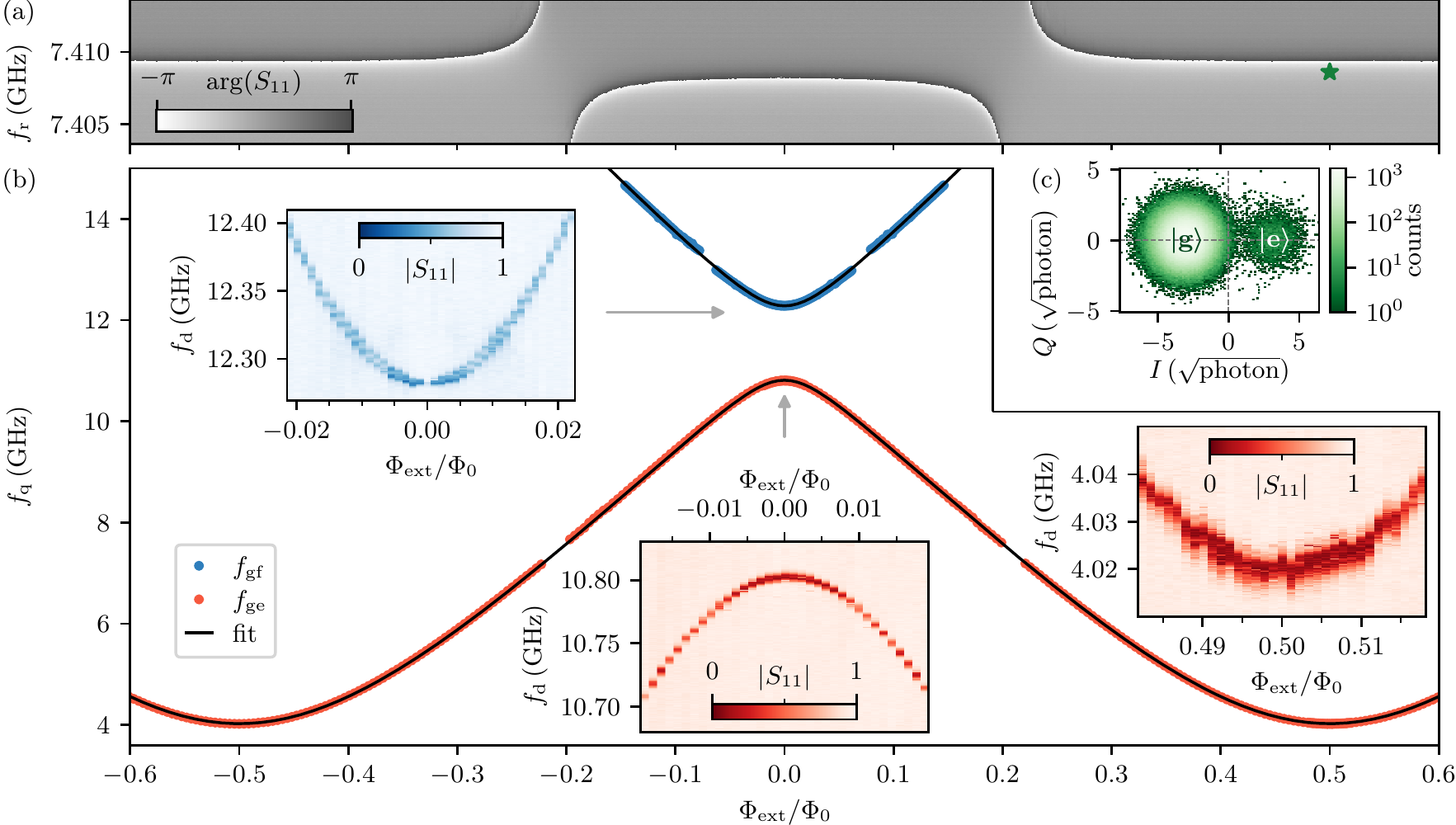}
\caption{\textbf{Gralmonium spectroscopy vs. external flux.} \textbf{(a)} Readout resonator phase response meaured in reflection. The field sweep reveals avoided level crossings between the qubit and the resonator. \textbf{(b)} Gralmonium spectrum measured by two-tone spectroscopy. The data for the $\ket{\text{g}}\to\ket{\text{e}}$ (red markers) and $\ket{\text{g}}\to\ket{\text{f}}$ (blue markers) transitions is extracted from continuous wave monitoring of the resonator while applying a second drive tone at $f_\text{d}$. The gaps in the extracted qubit frequencies around $\SI{7.4}{\giga\hertz}$ and $\SI{13.1}{\giga\hertz}$ result from avoided level crossings with the readout resonator and the first superinductor mode, respectively. From a fit to the spectrum (black lines), we extract $L_\text{q} = \SI{285}{\nano\henry}$, $E_\text{J}/\text{h}= \SI{23,4}{\giga\hertz}$ and $C_\Sigma = C_\text{q} + C_\text{J} = \SI{1.26}{\femto\farad}$ ($E_\text{C}^{\Sigma}/\text{h}=\SI{15}{\giga\hertz}$) for the qubit parameters (cf. \figref{fig:sample}). The insets show raw spectroscopy data at the sweet spots with the color scale corresponding to the single-port reflection amplitude. At half-flux (bottom right inset), the qubit frequency toggles from trace to trace (on a timescale of minutes). Note that the visibility of the $\ket{\text{g}}\to\ket{\text{f}}$ transition vanishes in the vicinity of $\Phi_\text{ext}=0$ (top left inset), as expected from the fluxonium selection rules~\cite{somoroff2021millisecond}. \textbf{(c)}~IQ histogram of contiguous reflection coefficient measurements at $\Phi_\text{ext}/\Phi_0=0.5$ and $f_\text{r}=\SI{7.4086}{\giga\hertz}$ (green marker in~(a)). Each point is integrated for $\SI{784}{\nano\second}$ at $\bar{n}\approx 10$ circulating photons in the readout resonator. Both $\ket{\text{g}}$ and $\ket{\text{e}}$ states are visible, separated by a dispersive shift $\chi/2\pi = \SI{-1.7}{\mega\hertz}$ (see also \suppref{sec:supp:dispersiveshift}), and their populations correspond to \SI{37}{\milli\kelvin} effective temperature.
}
\label{fig:spectrum}
\end{figure*}

The first indication of a functioning qubit coupled to the resonator is the measurement of avoided level crossings versus external flux (\figrefadd{fig:spectrum}{a}), which repeat periodically (\suppref{sec:supp:periodicity}). By measuring spectroscopy of the gralmonium (\figref{fig:spectrum}), we confirm that it is accurately modeled by the standard fluxonium Hamiltonian~\cite{fluxonium},
\begin{align}
    H &= 4 E_\text{C}^\Sigma n^2 + \frac{1}{2}E_\text{L}\left(\varphi - 2\pi\frac{\Phi_\text{ext}}{\Phi_0}\right)^2 - E_\text{J}\cos\varphi\,,
    \label{eq:fluxoniumHamiltonian}
\end{align}
with the grAl nano-junction serving as an SIS JJ with sinusoidal current-phase relation and effective Josephson energy $E_\text{J}$ (cf. \figrefadd{fig:sample}{d}). The operators $n$ and $\varphi$ correspond to the number of Cooper pairs and phase difference across the junction, respectively, $\Phi_0=\text{h}/2\text{e}$ is the superconducting magnetic flux quantum, $E_\text{L}=(\Phi_0 / 2\pi)^2 / L_\text{q}$ the inductive energy and $\Phi_\text{ext}$ the external flux through the gralmonium loop. \revise{We quantify the agreement between the gralmonium spectrum and a sinusoidal current-phase relation in \suppref{sec:supp:noncosine}. In total, we measured 20 spectra of gralmonium devices consistent with the fluxonium Hamiltonian (cf. \eqref{eq:fluxoniumHamiltonian}) across 11 wafers (cf. \suppref{sec:supp:othersamples}).}

\begin{figure*}[t!]
\centering
\includegraphics[width=\textwidth]{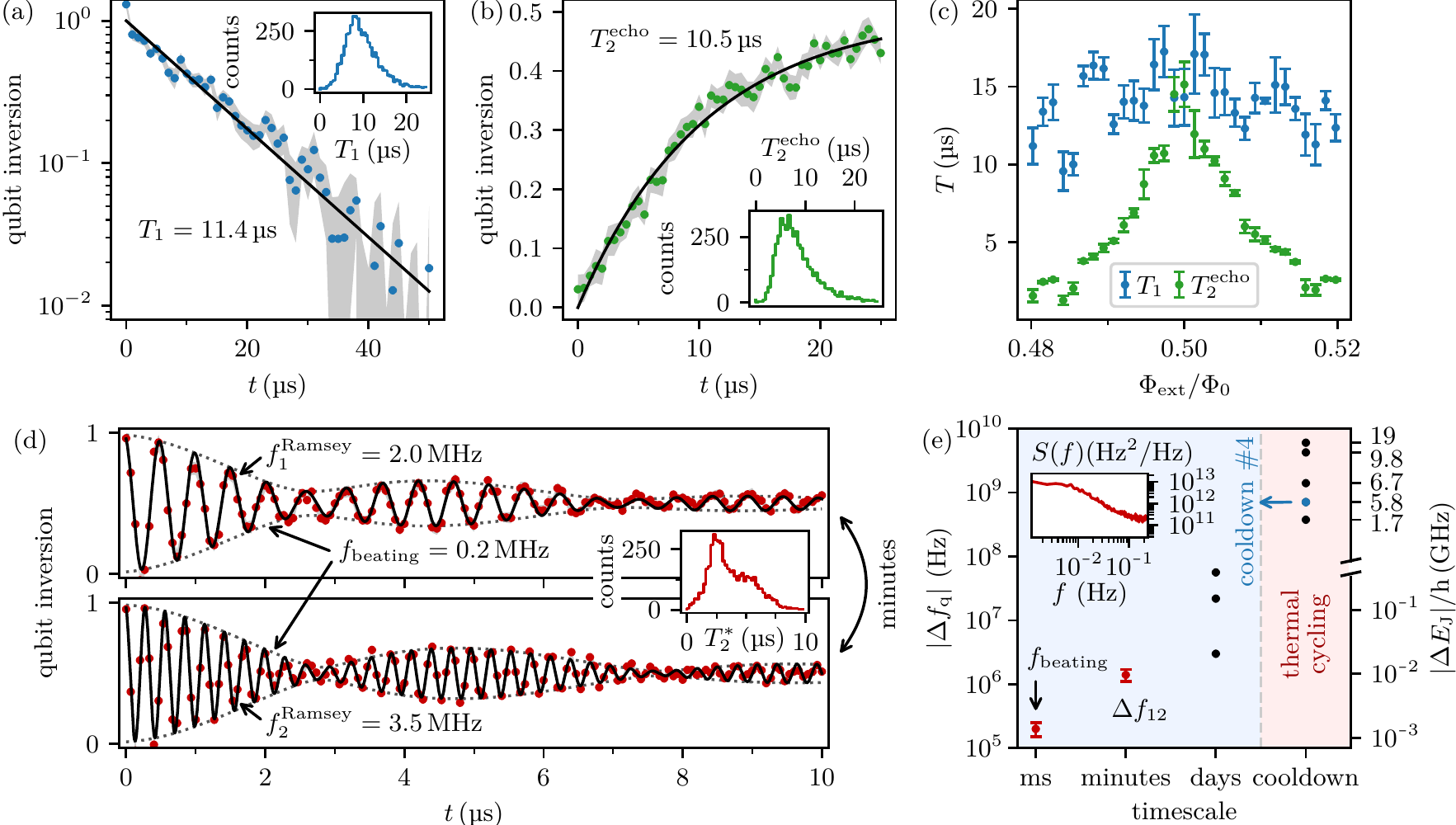}
\caption{\textbf{Time domain characterization of the gralmonium at half-flux bias.} \textbf{(a)}~Free decay energy relaxation and \textbf{(b)}~spin Hahn echo experiment with exponential fits (black lines) corresponding to $T_1=\SI{11.4}{\micro\second}$ and $T_2^\text{echo}=\SI{10.5}{\micro\second}$, respectively. \textbf{(c)}~$T_1$ and $T_2^\text{echo}$ measurements vs. external flux. The markers and errorbars represent the mean and standard error of the mean of decay times extracted in 5 repeated flux sweeps. \textbf{(d)}~Interestingly, Ramsey fringes measured with a nominal detuning of $\SI{2}{\mega\hertz}$, exhibit a beating pattern corresponding to two qubit frequencies separated by $f_\text{beating}=\SI{0.2}{\mega\hertz}$ (cf.~two-frequency fit in black and dotted envelope). Moreover, a comparison of the measurements shown in the top and bottom panel, which were acquired under identical conditions, illustrates jumps of the \revise{average qubit frequency by $\Delta f_{12}=\SI{1.5}{\mega\hertz}$}. We fit $T_2^*$ in the range of \SIrange{1}{10}{\micro\second} (inset). In panel \textbf{(e)} we summarize the different timescales on which we observe changes in the qubit frequency. As visible in (d), the toggling of the qubit frequency by $f_\text{beating}=\SI{0.2}{\mega\hertz}$ occurs on a timescale of milliseconds (Ramsey measurement time) and is accompanied by less frequent jumps of \revisex{$\Delta f_{12}^{Ramsey}=$}$\SI{1.5}{\mega\hertz}$ on a timescale of minutes. \revise{The inset shows the corresponding Lorentzian power spectrum (cf. \suppref{sec:supp:PSD})}. In addition, we report \SIrange{10}{100}{\mega\hertz} changes every few days during a cooldown. The largest frequency shifts are observed after thermal cycling to room temperature. The right axis shows the corresponding change of the Josephson energy $E_\text{J}$, which we identify as the cause of the qubit frequency changes. Note, that the measurements in panels (a)--(d) are obtained using 20 acquisitions of 100 averaged single shot qubit measurements; the error bands (grey) show the standard error of the mean over the 20 iterations and the results of 2000 individual iterations are histogrammed in the insets.
}
\label{fig:timedomain}
\end{figure*}

 \Figrefadd{fig:spectrum}{b} shows \revise{a} gralmonium spectrum up to \SI{14}{\giga\hertz}, measured by probing the readout resonator while applying a second microwave tone varying in frequency. A joint numerical fit~\cite{smithDiagonalization} of \eqref{eq:fluxoniumHamiltonian} to the $\ket{\text{g}}\to\ket{\text{e}}$ (red) and $\ket{\text{g}}\to\ket{\text{f}}$ (blue) transitions matches the data and gives $E_\text{J}=\SI{23.4}{\giga\hertz}$, in agreement with the expected range for the dimension of the grAl nano-junction~\cite{felix,graltransmon}. The fitted capacitance across the junction $C_\Sigma=C_\text{q} + C_\text{J}$ is $\SI{1.26}{\femto\farad}$. The $\ket{\text{g}}\to\ket{\text{f}}$ transition is suppressed around zero flux (\figrefadd{fig:spectrum}{b}, top left inset), following the fluxonium selection rules~\cite{fluxonium}. \revise{Additionally, due to large quantum fluctuations facilitated by the small value of the nano-junction capacitance, even at zero flux the $\ket{g}$ and $\ket{e}$ eigenfunctions of the gralmonium are delocalized (cf.~\suppref{sec:supp:ZeroFluxTD}).} Surprisingly, the $\ket{\text{g}}\to\ket{\text{e}}$ transition linewidth does not narrow as the flux bias is tuned towards the sweet spots, \revise{foreshadowing the presence of an additional decoherence mechanism besides flux noise, namely critical current fluctuations.} On a timescale of\revisex{ single} \revise{a few} traces (minutes), a toggling of the half-flux frequency is observed (bottom right inset), which will be discussed in more detail in \figref{fig:timedomain}. In \figrefadd{fig:spectrum}{c} we plot an IQ histogram of the measured reflection coefficient at half-flux bias. The two distributions visible in the plot correspond to the steady state populations of $\ket{g}$ and $\ket{e}$.

\vspace{2mm}

We complete the gralmonium characterization with time domain measurements at half-flux. The top row of \figref{fig:timedomain} shows free decay energy relaxation and spin Hahn echo measurements with single exponential decay times on the order of \SI{10}{\micro\second}. The maximum $T_2^\text{echo}$ is reached at the $\Phi_\text{ext}/\Phi_0=0.5$ sweet spot, where the spectrum is first order insensitive to flux noise (cf.~\figrefadd{fig:timedomain}{c}) \revise{and we discuss the decoherence budget in \suppref{sec:supp:T2vsFlux}}. The $T_1$ values extracted from free decay and from quantum jump traces (see \suppref{sec:supp:quantumjumps}) are comparable, which is indicative of a photon number independent energy relaxation, as demonstrated for other grAl fluxonium qubits~\cite{nbarfluxonium}.

Departing from the behavior of standard fluxonium qubits using mesoscopic SIS junctions, we observe conspicuous fluctuations of the nano-junction $E_\text{J}$ in gralmonium devices. The bottom row of \figref{fig:timedomain} summarizes the corresponding changes of the qubit frequency on different timescales. On a timescale faster than single measurements ($\approx\si{\milli\second}$), the qubit toggles between two frequencies  $f_\text{beating}=\SI{0.2}{\mega\hertz}$ apart, resulting in Ramsey fringes with a beating pattern (\figrefadd{fig:timedomain}{d}). In addition, on a timescale of minutes, we observe frequency jumps \revise{of $\SI{1.5}{\mega\hertz}$, visible both in Ramsey fringes (\figrefadd{fig:timedomain}{d}) and in continuous wave spectroscopy (cf.~\figref{fig:spectrum} and \suppref{sec:supp:splitspectrum})}. \revise{This finding is supported by a Lorentzian power spectrum characteristic for random telegraphic noise with switching rate $\Gamma_\text{RTN}=\SI{9.4}{\milli\hertz}$ (cf. \suppref{sec:supp:PSD})}. Moreover, we observe \SIrange{10}{100}{\mega\hertz} jumps every few days while the sample remains at cryogenic temperature and the largest changes occur after thermal cycling (\figrefadd{fig:timedomain}{e}).

These fluctuations are not entirely surprising, if one recalls that the first fluxonium levels at half-flux are determined by tunnelling through the Josephson barrier. The frequency of the first transition can be approximated by the phase slip rate~\cite{phasesliprate}
\begin{equation}
    \nu= \frac{4}{\sqrt{\pi}} \left(8 E_\text{J}^3 E_\text{C}^\Sigma\right)^{1/4} e^{-\sqrt{8E_\text{J}/E_\text{C}^\Sigma}} \, .
    \label{eq:phasesliprate}
\end{equation}

Consequently, the qubit frequency is exponentially sensitive to $E_\text{J}/ E_\text{C}^\Sigma$, rendering the gralmonium susceptible to microscopic changes in the $(\SI{20}{\nano\meter})^3$ volume of the grAl nano-junction and its close vicinity.\revisex{ The list of possible culprits~\cite{Siddiqi2021} includes, but is not limited to charge noise, vacancies or interstitial impurities, positional defects and paramagnetic defects. While we do find correlated changes of the intrinsic nano-junction $E_\text{C}$ and $E_\text{J}$ ($\text{\suppref{sec:supp:paramtersvscd}}$),} The spectrum is more sensitive to $E_\text{J}$ changes (right hand axis of \figrefadd{fig:timedomain}{e} and \suppref{sec:supp:splitspectrum}) because $E_\text{C}^\Sigma$ is bounded by the \revise{value of the} interdigitated capacitance $C_\text{q}\approx\SI{0.8}{\femto\farad}$\revise{, obtained from finite element simulations. The fact that $E_\text{J}$ and $E_\text{C}$ fluctuations appear correlated (cf. \suppref{sec:supp:paramtersvscd}) is an indication that the quasiparticle capacitance of the nano-junction \cite{Eckern1984}, which we estimate in the range of \SI{0.05}{\femto\farad}, plays a visible role in the gralmonium.}

\revise{The list of possible culprits for the intrinsic nano-junction fluctuations includes, but is not limited to (i)~structural changes, i.e. tunneling crystalline defects, vacancies, interstitial impurities or adsorbed molecules~\cite{Siddiqi2021}, (ii)~charge noise due to changes in locally trapped charges via the Aharonov--Casher effect \cite{Friedman2002, Pop2012Aharonov, Manucharyan2012Aharonov} and (iii)~paramagnetic defects. Future experiments to discriminate between these candidates could involve a local electric field bias using a gate electrode, applying mechanical stress \cite{Lisenfeld2019} or in-plane magnetic field. Moreover, the fluctuations might be reduced in future devices by using cold substrate deposition, which has been shown to yield smaller and more regular grains \cite{grainsize}, or post processing such as hydrogen or laser annealing \cite{Hertzberg2021}. Beyond detecting changes in its microscopic structure, the nano-junction in a gralmonium can also be used as a local probe of magnetic field, such as spins in the substrate below the junction, or spin qubits in semiconductors.}

In conclusion, we have demonstrated that a grAl nano-junction can provide the source of nonlinearity in a superconducting fluxonium qubit with $\approx \SI{10}{\micro\second}$ coherence, enabling its fabrication without the use of mesoscopic JJs, in a single layer of zero angle deposited grAl.  Spectroscopy confirms that the gralmonium is governed by the standard fluxonium Hamiltonian with a Josephson energy of the grAl nano-junction on the order of conventional qubit JJs. In contrast to mesoscopic SIS JJs, the intrinsically small capacitance of the nano-junction, $C_\text{J}<\SI{1}{\femto\farad}$, opens a new parameter regime, particularly relevant for high impedance circuits where large quantum fluctuations of the phase are desirable~\cite{Groszkowski2018Apr, zeropihouck}. \revise{Notably, devices such as the 0-$\pi$ qubit biased at half flux bias would not suffer from $E_\text{J}$ fluctuations of the nano-junctions. Moreover, nano-junctions in which flux tunneling is suppressed (cf. \suppref{sec:supp:periodicity}) can be used to replace the overlap JJs in SNAIL elements \cite{frattini20173,Grimm2020} for parametric devices.}

\revise{Beyond grAl}, the nano-junction concept presented here can probably be implemented using other granular superconductors\revise{~\cite{Beloborodov2007Apr} or}\revisex{ It might also possible to implement the gralmonium circuit using} homogeneously disordered superconductors close to the superconducting to insulating transition, where the spatial inhomogeneity of the gap creates a structure reminiscent of granular superconductivity~\cite{sacepe}. 

The nanoscopic footprint of the nonlinear element in the gralmonium, combined with \si{\micro\second} coherence times, provides an exciting resource for seemingly antithetic reasons. On one hand, the \textit{reduced} susceptibility to external magnetic fields enables utilization of the grAl nano-junction in hybrid architectures employing magnetic fields. On the other hand, the \textit{increased} susceptibility of the nano-junction to microscopic defects and noise channels in its immediate vicinity serves as a sensitive detector and offers a new handle for their characterization. These results open a window of opportunity for material science to directly impact the development of coherent superconducting hardware \revise{with} the corresponding nano-junction stability \revise{providing} an unambiguous metric for improvement.

\unrevise{\vspace{0.7cm}}

\revise{
\section*{Data Availability}
All relevant data are available from the corresponding author upon reasonable request.
}

\section*{Acknowledgements}
We are grateful to U. Vool for fruitful discussions and we acknowledge technical support from S. Diewald, A. Eberhardt, M.K. Gamer, A. Lukashenko and L. Radtke. Funding is provided by the Alexander von Humboldt foundation in the framework of a Sofja Kovalevskaja award endowed by the German Federal Ministry of Education and Research and by the European Union’s Horizon 2020 programme under No.~899561 (AVaQus). P.P. and I.M.P. acknowledge support from the German Ministry of Education and Research (BMBF) within the QUANTERA project SiUCs (FKZ: 13N15209). D.R., S.G., P.W. and W.W. acknowledge support from the European Research Council advanced grant MoQuOS (No.~741276). Facilities use was supported by the Karlsruhe Nano Micro Facility (KNMFi) and KIT Nanostructure Service Laboratory (NSL). We acknowledge qKit for providing a convenient measurement software framework.

\bibliography{references}

\balancecolsandclearpage

\onecolumngrid
\section*{Appendices}
\vspace{0.6cm}
\twocolumngrid
\appendix

\section{Fabrication Details}
\label{sec:supp:fabrication}
All samples discussed in this manuscript are fabricated on c-plane, double side polished sapphire substrates using lift-off e-beam lithography. A bi-layer resist stack of \SIrange{700}{800}{\nano\meter} MMA EL-13 and \SI{300}{\nano\meter} PMMA A4 and a \SI{10}{\nano\meter} chromium anti-static layer is used for writing with a \SI{100}{\kilo\electronvolt} e-beam writer. The structures are developed in a spray developer with a MIBK-isopropanol mixture with volume ratio 1:3. Before the metal deposition in a PreVac  evaporation system, the substrate is cleaned with a Kaufmann ion source in an Ar/O$_2$ descum process and the vacuum is improved using titanium gettering. The \SI{20}{\nano\meter} thick granular aluminum film is deposited under zero angle, at room temperature and a deposition rate around $\approx\SI{1}{\nano\meter\per\second}$ in a dynamic oxygen atmosphere resulting in a chamber pressure of $10^{-5}-10^{-4}\,\si{\milli\bar}$. The sheet resistance for the main text sample is $\SI{1.5}{\kilo\ohm/\sq}$.

\section{Sample Holder}
\label{sec:supp:sampleholder}
\begin{figure}[]
\centering
\includegraphics[width=\columnwidth]{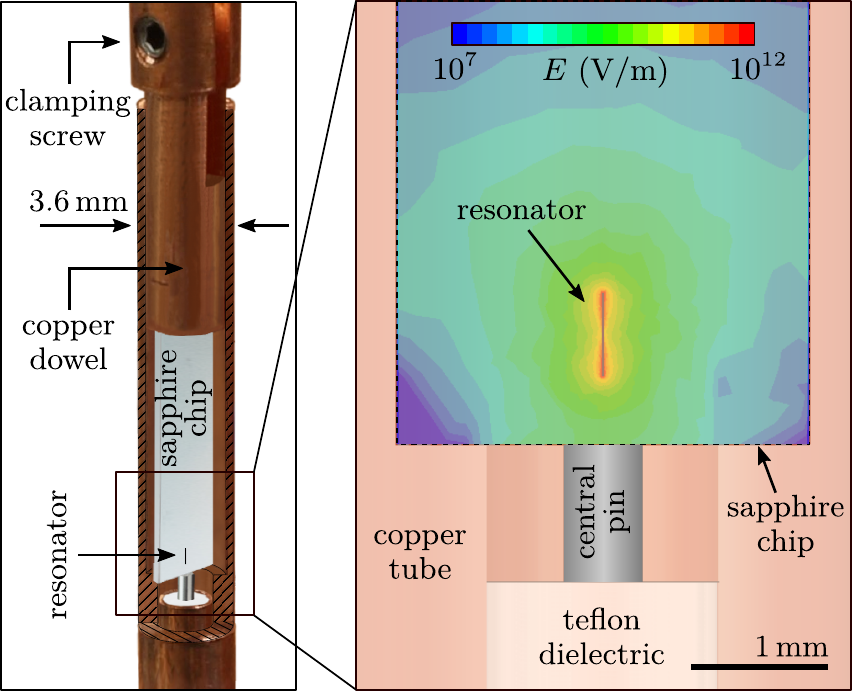}
\caption{
\textbf{Cylindrical waveguide sample holder (left side) and color plot of a finite element simulation of the electric field distribution (right side).} The cylindrical waveguide design is the same \revise{we used in Ref.~\cite{kiril} to measure grAl resonators in magnetic fields up to \SI{1}{\tesla}. The sapphire chip is fixed by a copper dowel, which is tightened against the walls of the waveguide copper tube using the clamping screw. The readout resonator is located at a distance of about \SI{0.5}{\milli\meter} from the bottom edge of the chip, close to the stripped central pin of a \SI{2.2}{\milli\meter} coaxial cable with teflon dielectric.} The electric field scale corresponds to an energy of \SI{1}{\joule}.
}
\label{fig:supp:sampleholder}
\end{figure}
In \figref{fig:supp:sampleholder} we show the copper sample holder used for cryogenic microwave reflection measurements (cf. also Ref.~\cite{kiril}). The \SI{3}{\milli\meter} inner diameter of the waveguide corresponds to a cut-off frequency of \SI{60}{\giga\hertz}. The microwave coupling of the readout resonator is given by evanescent waves from the central pin of the coaxial cable connected to the waveguide. 

\section{Qubit-Resonator Avoided Crossings}
\label{sec:supp:periodicity}
\begin{figure}[tb]
\centering
\includegraphics[width=\columnwidth]{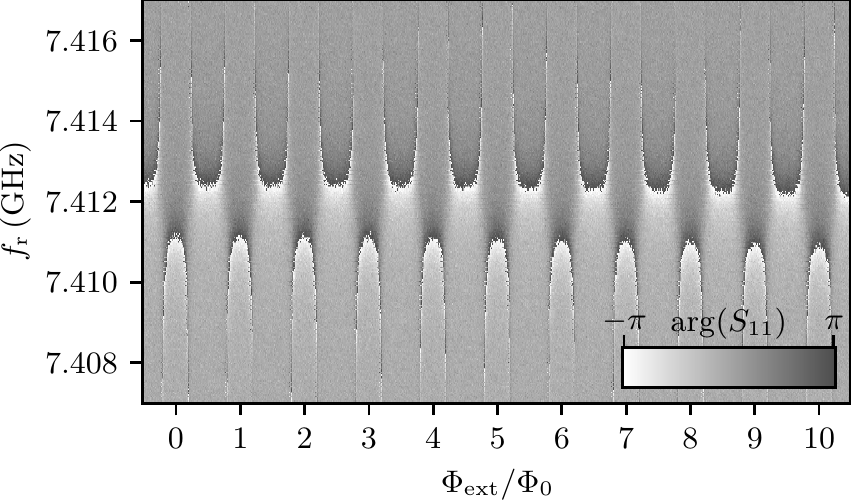}
\caption{
\textbf{Extended flux sweep of the readout resonator up to $\Phi_\text{ext}/\Phi_0=10$.} The strict periodicity is superimposed by a $\approx\SI{200}{\kilo\hertz}$ parabolic frequency shift due to screening currents in the antenna. The measurement was performed in a previous cooldown compared to \figrefadd{fig:spectrum}{a}, which is why the resonance frequency is \SI{2.5}{\mega\hertz} higher.
}
\label{fig:supp:periodicity}
\end{figure}
\begin{figure}[bt]
\centering
\includegraphics[width=\columnwidth]{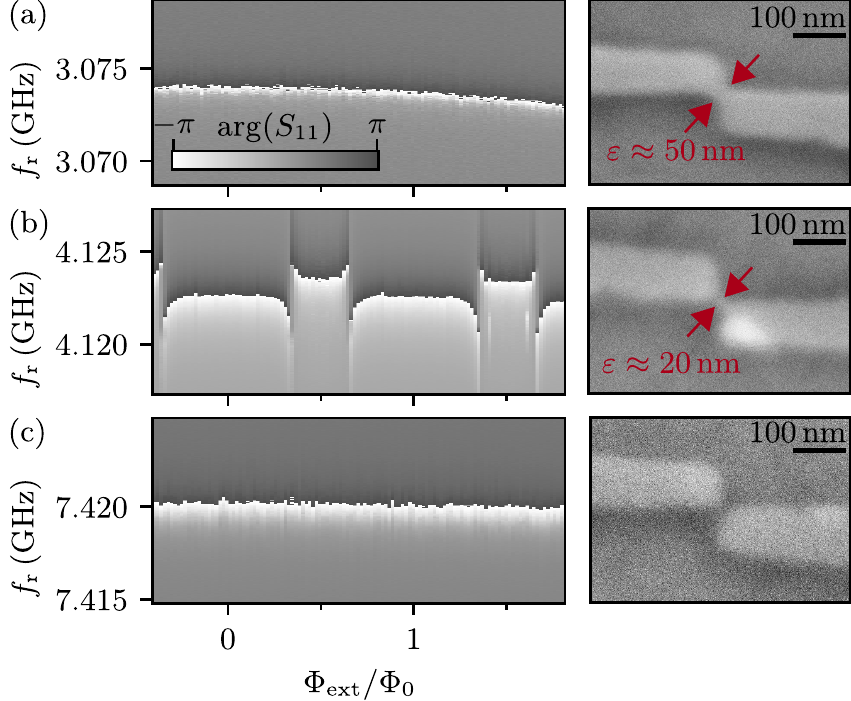}
\caption{\textbf{Comparison of grAl nano-junction regimes.} The left column shows the phase of the reflection coefficient measured versus external flux for three different samples \revise{with nominally identical gralmonium and nano-junction design}. The right column depicts the corresponding SEM image of the grAl nano-junction, taken after the cooldown. \textbf{(a)} The resonator frequency slightly decreases with increasing external flux due to a too wide nano-junction ($\varepsilon\approx\SI{50}{\nano\meter}$) and resulting screening currents in the loop. \textbf{(b)} The flux sweep shows periodic avoided level-crossings, which are the signature of a functioning gralmonium attached to the resonator. The nonlinear element is in the same regime of $\varepsilon\approx\SI{20}{\nano\meter}$ as the main sample (\figref{fig:sample}). \textbf{(c)} The resonator frequency is constant in this field range because the nano-junction is interrupted.
}
\label{fig:supp:nocross}
\end{figure}
The signature of a gralmonium device operating in the desired frequency range and coupled to the readout resonator is the measurement of qubit-resonator avoided level crossings (cf. \figrefadd{fig:spectrum}{a}), periodic in flux. In \figref{fig:supp:periodicity} we present measurements of the qubit-resonator anti-crossings for the main text device, extending over 10 $\Phi_0$ periods. The \SI{200}{\kilo\hertz} frequency shift observable between the anti-crossing patterns at $\Phi_\text{ext}/\Phi_0=0$ and $\Phi_\text{ext}/\Phi_0=10$, is due to the frequency dependence of the grAl antenna as a function of out-of-plane field, consistent with Ref.~\cite{kiril}. 

The $E_\text{J}/E_\text{C}$ ratio for the grAl nano-junction is highly sensitive to its width, which can lead to very different spectra for nominally identically fabricated devices, simply due to lithography variability. A rapid method to select working devices is to measure their field dependence. As an example, in \figref{fig:supp:nocross} we compare three devices \revise{ with nominally identical design for the gralmonium and nano-junction,} fabricated in the same evaporation on the same chip and measured in the same cooldown. The flux sweeps reveal three different junction regimes. The device presented in \figrefadd{fig:supp:nocross}{(a)} shows a \SI{1}{\mega\hertz} shift but no qubit-resonator avoided level crossings. This can be explained by the fact that the ratio $E_\text{J}/E_\text{C}$ for this nano-junction is much larger than unity, and quantum fluctuations of the phase are suppressed. In this case the flux bias induces persistent currents in the device loop\revise{, responsible for the \SI{1}{\mega\hertz} frequency shift (which is much larger than the frequency dependence vs. field of the bare antenna, as shown in \figrefadd{fig:supp:nocross}{(c)})}. The corresponding SEM micrograph of the nano-junction confirms its relatively large size, with $\varepsilon \approx \SI{50}{\nano\metre}$. The flux sweep presented in \figrefadd{fig:supp:nocross}{(b)} shows qubit-resonator avoided level crossings similar to \figrefadd{fig:spectrum}{a}, and the corresponding SEM micrograph confirms the smaller width of the grAl nano-junction of this device compared to \figrefadd{fig:supp:nocross}{(a)}. In \figrefadd{fig:supp:nocross}{(c)} we show the measurement of a device where the grAl nano-junction is interrupted. As expected for this flux range, we do not observe a change in the frequency of the readout resonator.

\revise{Based on SEM imaging of 53 devices, we find 22\% of nano-junctions to be in the regime shown in \figrefadd{fig:supp:nocross}{b}, i.e. connected junctions with $\varepsilon \leq \SI{25}{\nano\meter}$. These} statistics can \revisex{potentially} be improved by using thinner resist layers \revise{ and by replacing the chromium anti-static coating for e-beam writing (cf.~\suppref{sec:supp:fabrication}) with aluminum, which yields reduced roughness. Thanks to the single-layer circuit design, etching is a tantalizing alternative  to lift-off fabrication with the added benefit of using sub \SI{10}{\nano\meter} resolution negative \mbox{e-beam} resists. In addition to improving the lithography reliability, first experiments indicate that post-processing samples with wet-etching or annealing can tune nano-junctions closer to a desired $E_\text{J}$ regime.} Moreover, we expect that nano-junctions as shown in \figrefadd{fig:supp:nocross}{(a)} might also lead to working gralmonium devices for sufficiently large sheet resistivity, as long as the grAl film remains superconducting.

\section{Extraction of Dispersive Shift}
\begin{figure}[t]
\centering
\includegraphics[width=\columnwidth]{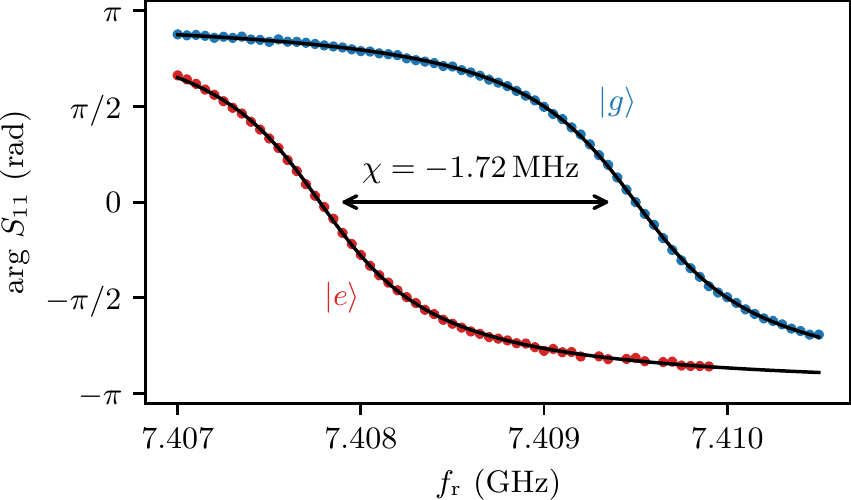}
\caption{
\textbf{Dressed resonator phase responses for qubit in $\ket{g}$ (blue) and $\ket{e}$ (red).} The data is extracted from IQ histograms similar to the one shown in \figrefadd{fig:spectrum}{c} with a $\pi/2$ qubit pulse applied before the readout pulse, in order to balance the populations of the qubit states. By fitting the phase responses, we extract a dispersive shift of $\chi/2\pi=\SI{-1.72}{\mega\hertz}$ and a resonator linewidth of $\kappa/2\pi=\SI{1.00}{\mega\hertz}$.
}
\label{fig:supp:dispersiveshift}
\end{figure}
\label{sec:supp:dispersiveshift}
In \figref{fig:supp:dispersiveshift}, we show the method to extract the dispersive shift between the dressed resonator responses for the qubit in $\ket{g}$ and $\ket{e}$. The regime of $|\chi|>\kappa$ enables qubit readout with maximum \SI{180}{\degree} phase separation (\figref{fig:spectrum}).

\revise{
\section{Sinusoidal Current Phase Relation of the Nano-Junction}
\label{sec:supp:noncosine}
\begin{figure*}[t]
\centering
\includegraphics[width=\textwidth]{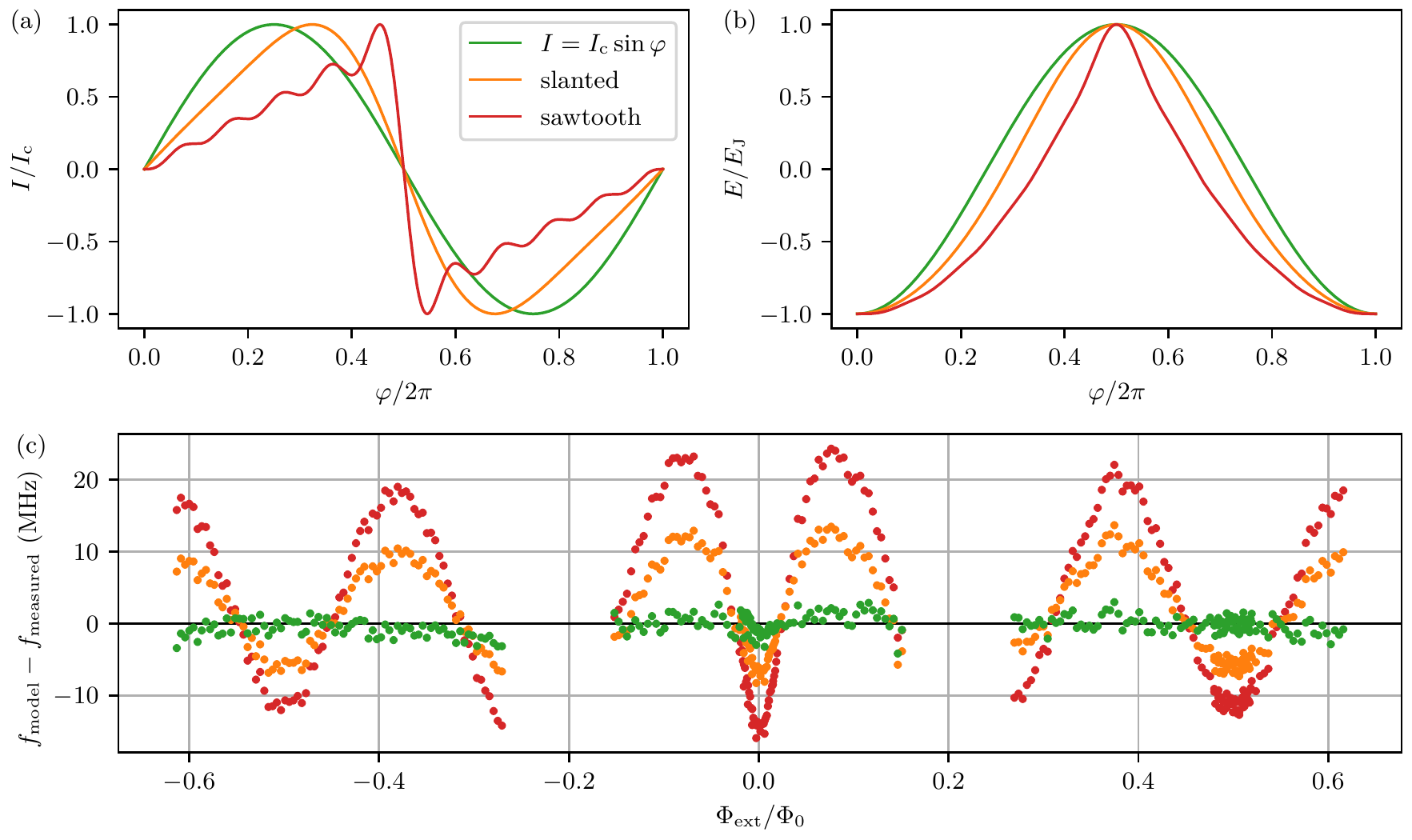}
\caption{\revise{
\textbf{Comparison between the measured spectrum and the fluxonium Hamiltonian.} \textbf{(a)} Current phase relations (C$\varphi$Rs) used for modeling. In addition to the standard SIS JJ sinusoidal C$\varphi$R, $I=I_\text{c}\sin\varphi$ (green), we consider two other models containing higher order Josephson harmonics \cite{GolubovCphiR}: a slanted C$\varphi$R, $I=I_\text{c}(\sin\varphi-0.25\sin 2\varphi+0.05\sin 3\varphi)$ (orange), and a sawtooth-like C$\varphi$R, $I=I_\text{c}\sum_{n=1}^{10} (-1)^{(n+1)} \sin(n\varphi)/n$ (red). \textbf{(b)} Energy phase relations corresponding to the C$\varphi$Rs in (a). In \textbf{(c)} we plot the difference between the measured eigenfrequencies $f_\text{measured}$ for the $\ket{g}\to\ket{e}$ transition (cf. \figref{fig:spectrum}) and the numerical diagonalization $f_\text{model}$ using Hamiltonian \eqref{eq:fluxoniumHamiltonian}, where the Josephson term is given by the energy phase relations plotted in (b). We exclude a $\pm\SI{1}{\giga\hertz}$ interval in the vicinity of the avoided level crossing with the resonator ($|\Phi_\text{ext}/\Phi_0|\approx 0.2$). Notably, the standard sinusoidal C$\varphi$R used in the main text (green) shows no deviations within $\pm\SI{2}{\mega\hertz}$, which is the resolution of the measurement. In contrast, the non-sinusoidal C$\varphi$Rs (orange and red corresponding to (b)) show order of magnitude larger and systematic deviations. Based on this measurement, we place an upper bound of \SI{5}{\percent} for higher harmonics contributions in the C$\varphi$R of the nano-junction.
}}
\label{fig:supp:noncosine}
\end{figure*}
In the following, we quantify how accurately the standard fluxonium Hamiltonian \eqref{eq:fluxoniumHamiltonian} with sinusoidal nano-junction current-phase relation (C$\varphi$R) describes the measured gralmonium spectrum shown in \figref{fig:spectrum} in the main text. To do so, in addition to numerically diagonalizing \eqref{eq:fluxoniumHamiltonian} using the pure sinusoidal C$\varphi$R we also consider two other C$\varphi$Rs increasingly deviating from a pure sine. We construct these C$\varphi$Rs by adding higher order terms $\sin(n\varphi)\,(n>1)$ \cite{GolubovCphiR}. The diagonalization is performed using a straightforward extension of the same numerical method used in the main text \cite{smithDiagonalization}. We fit the model to the experimental data for each C$\varphi$R individually. For the purely sinusoidal case it is feasible to also include the coupling to the readout resonator in the model in order to describe the avoided-level crossings between qubit and resonator, while for the non-sinusoidal C$\varphi$Rs this task becomes computationally intensive.

The C$\varphi$Rs used for the analysis, their corresponding energy phase relations and the comparison between model and measurement are shown in \figref{fig:supp:noncosine}. Remarkably, the standard sinusoidal C$\varphi$R matches the data to within $\pm\SI{2}{\mega\hertz}$. In contrast, the models with higher order contributions systematically deviate from the data by an order of magnitude. Based on the spread of the measured values for the $\ket{g}\to\ket{e}$ frequencies, we cannot rule out higher order Josephson harmonics with a relative contribution smaller than \SI{5}{\percent}.
}

\revise{
\section{Zero flux coherence}
\label{sec:supp:ZeroFluxTD}
\begin{figure*}[bt]
\centering
\includegraphics[width=\textwidth]{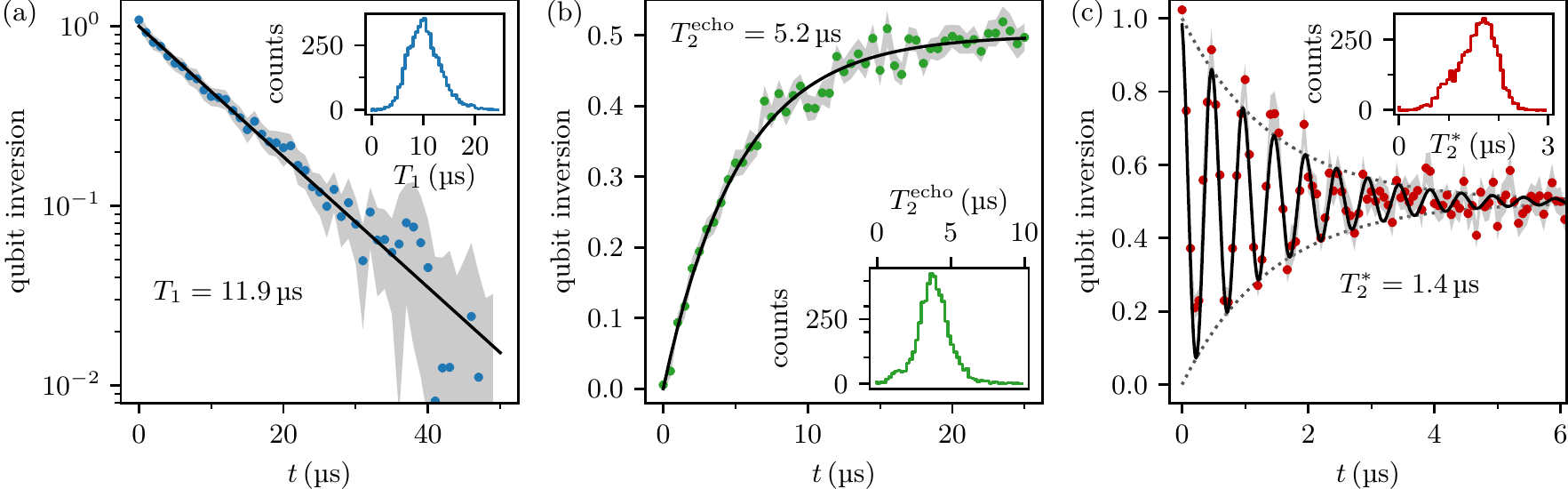}
\caption{\revise{
\textbf{Time domain characterization at zero flux.} \textbf{(a)}~Free decay energy relaxation and \textbf{(b)}~spin Hahn echo experiment with exponential fits (black lines). While the average $T_1$ is comparable to half flux, $T_2^\text{echo}$ is a factor of 2 lower.  \textbf{(c)} Ramsey fringes with a nominal detuning of \SI{2}{\mega\hertz}. In contrast to half flux, we do not observe a revival of Ramsey fringes at zero flux and the fit (black line) consists of a single sine wave with exponential envelope (dotted lines).
}}
\label{fig:supp:zerofluxTD}
\end{figure*}
\begin{figure}[t]
\centering
\includegraphics[width=\columnwidth]{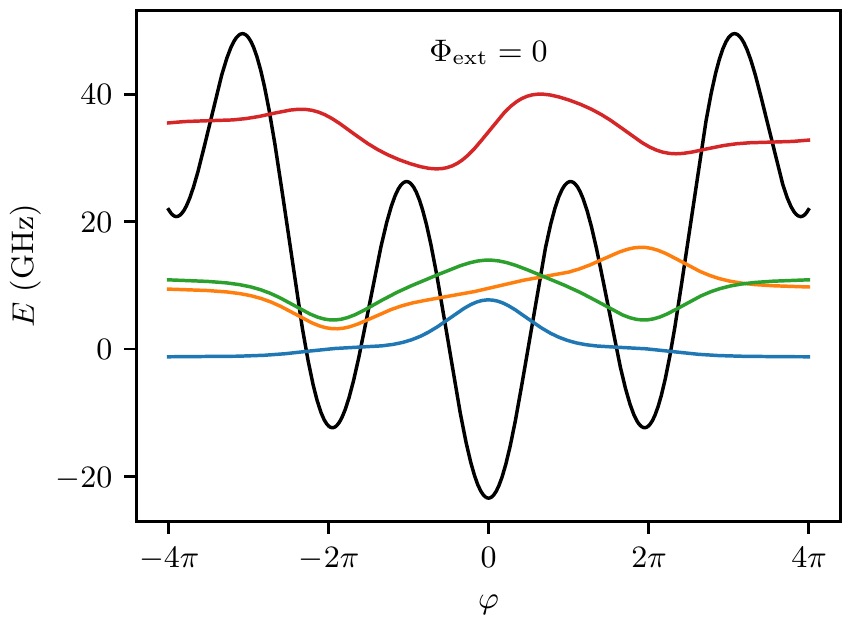}
\caption{\revise{
\textbf{Gralmonium energy potential (black) and wavefunctions (colored) at zero flux bias.}
We plot the first four wavefunctions, vertically offset by their eigenenergy. Note that all wavefunctions are significantly delocalized over three potential wells and, for this reason, the $\ket{g}\to\ket{e}$ transition is not plasmon-like.
}}
\label{fig:supp:potential}
\end{figure}

In \figref{fig:supp:zerofluxTD} we present energy relaxation and coherence measurements of the $\ket{g}\to\ket{e}$ transition performed at zero flux bias. Notably, the transition is not plasmon-like in our device, which is illustrated in \figref{fig:supp:potential} by the energy potential and wavefunctions.
}

\revise{
\section{Gralmonium Decoherence Around Half Flux}
\label{sec:supp:T2vsFlux}
\begin{figure}[t]
\centering
\includegraphics[width=\columnwidth]{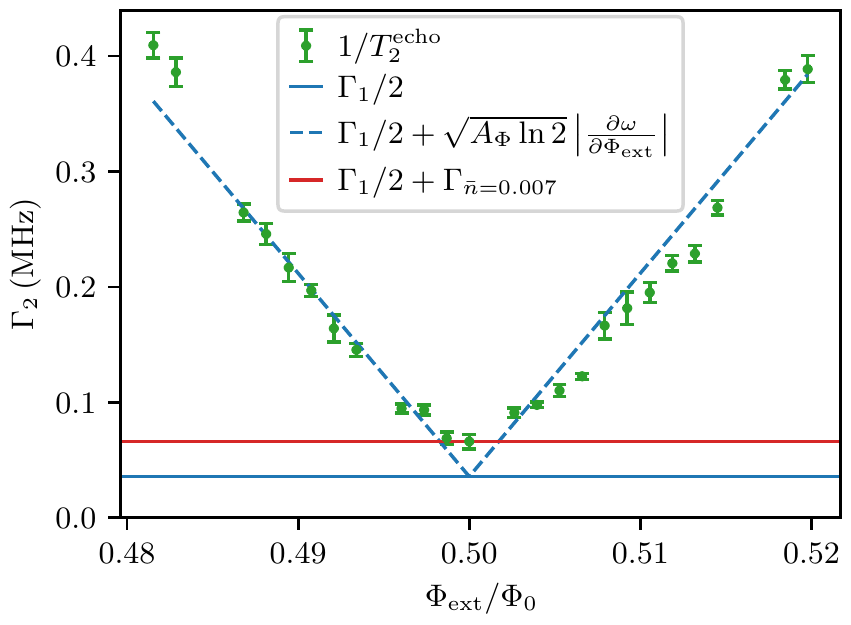}
\caption{\revise{
\textbf{Analysis of gralmonium decoherence around half flux bias.} Green markers show the echo decoherence rate $\Gamma_2^\text{echo}=1/T_2^\text{echo}$ based on the same data as shown in \figrefadd{fig:timedomain}{c} in the main text. The contribution from energy relaxation $\Gamma_1/2$ is indicated by the blue horizontal line. From the linear increase of $\Gamma_2^\text{echo}$ we extract a flux noise amplitude of $A_\Phi=\SI{30}{\micro\Phi_0}$ (dashed blue lines). The residual decoherence exactly at half flux is consistent with shot noise corresponding to an average thermal population of $\bar{n}\approx 0.007$ in the readout resonator, but could also stem from higher order flux noise.
}}
\label{fig:supp:T2vsflux}
\end{figure}
\begin{figure*}[t]
\centering
\includegraphics[width=\textwidth]{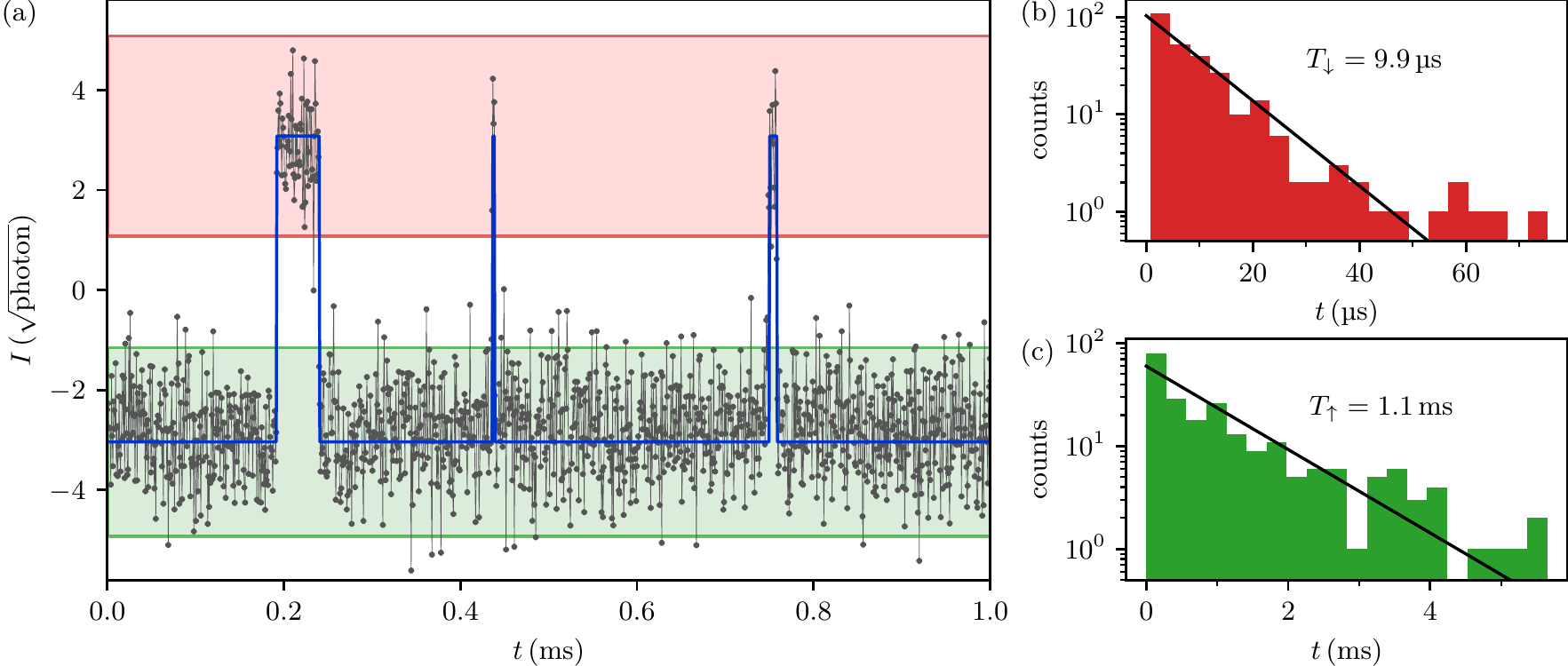}
\caption{
\textbf{Quantum jump measurement at half-flux bias point of the gralmonium.} \textbf{(a)} Example time trace of the in-phase component $I$ of the readout resonator reflection coefficient (black connected markers). We apply a continuous readout tone, populating the resonator with $\bar{n}\approx 10$ photons, and we demodulate contiguous windows of $\tau = \SI{784}{\nano\second}$. The qubit state is assigned using a latching filter (blue line) based on the $\pm2\sigma$ bands (green and red for $\ket{g}$ and $\ket{e}$, respectively) around the mean values of the qubit states. The corresponding IQ histogram of the demodulated data for a total measurement time of \SI{500}{\milli\second} is shown in \figrefadd{fig:spectrum}{c}. In \textbf{(b)} and \textbf{(c)} we histogram the durations spent by the qubit in the excited and ground state, respectively. Maximum likelihood exponential fits yield the average relaxation and excitation times $T_{\downarrow}=\SI{9.9}{\micro\second}$ and $T_{\uparrow}=\SI{1.1}{\milli\second}$, respectively.
}
\label{fig:supp:quantumjumps}
\end{figure*}
Here, we discuss in more detail the $T_2^\text{echo}$ data in the vicinity of $\Phi_\text{ext}/\Phi_0=0.5$ shown in \figrefadd{fig:timedomain}{c} in the main text. The corresponding decoherence rate $\Gamma_2^\text{echo}=1/T_2^\text{echo}$ is plotted in \figref{fig:supp:T2vsflux} (green markers). The decoherence budget consists of energy relaxation $\Gamma_1/2$, flux noise  $\Gamma_{\Phi}(\Phi_\text{ext})$, photon shot noise due to stray photons in the resonator $\Gamma_{\bar{n}}$ and critical current noise in the nano-junction and superinductor $\Gamma_{I_\text{c}}$:
\begin{equation}
    \Gamma_2^\text{echo} = \frac{\Gamma_1}{2} + \Gamma_{\Phi}(\Phi_\text{ext}) + \Gamma_{\bar{n}} + \Gamma_{I_\text{c}}\,.
\end{equation}
Energy relaxation vs. external flux is flat in the measured range (cf. \figrefadd{fig:timedomain}{c}) with an average $T_1=\Gamma_1 ^{-1}=\SI{14}{\micro\second}$, which corresponds to a constant contribution to decoherence $\Gamma_1 / 2$ indicated by the horizontal blue line in \figref{fig:supp:T2vsflux}. We observe a linear increase of $\Gamma_2^\text{echo}$ around half flux bias and from the slope we extract a $1/f$ flux noise amplitude of $A_\Phi=\SI{30}{\micro\Phi_0}$. Note that due to limited resolution and SNR of the echo measurements we cannot exclude an additional, flux dependent, Gaussian component in the decay curves, which is why we use an exponential to consistently fit the entire data set. Exactly at half flux, $\Gamma_2^\text{echo}$ remains approximately a factor of 2 above the $T_1$-limit $\Gamma_1/2$. This residual dephasing could be either due to higher order flux noise contributions or photon shot noise. The dephasing due to photon shot noise in the low photon limit is given by \cite{Zhang2017, Clerk2007}
\begin{align}
   \Gamma_{\bar{n}} = \frac{\bar{n} \kappa \chi^2}{\kappa^2 + \chi^2} \,.
\end{align}
Based on the independently measured values of $\chi$ and $\kappa$ (cf. \suppref{sec:supp:dispersiveshift}), the residual dephasing rate corresponds to a thermal population of $\bar{n}\approx 0.007$, consistent with literature (see e.g.~\cite{Zhang2017}). Finally, the critical current noise is the only contribution which can be filtered efficiently by the echo sequence, and is likely explaining the factor of 3 higher $T_2^\text{echo}$ compared to Ramsey decay time $T_2^*$ at half flux. The echo sequence filters noise on timescales between \SI{10}{\micro\second} (the filter function cut-off) and \si{\milli\second} which is already captured by the two frequency fit for the Ramsey measurement.
}

\section{Quantum Jump Analysis}
\label{sec:supp:quantumjumps}
In \figref{fig:supp:quantumjumps} we analyze the contiguously measured reflection coefficient of the readout resonator as a function of time. The measured data is the same as histogrammed in \figrefadd{fig:timedomain}{c}. We rotated the IQ plane such that the qubit state information is encoded in the in-phase quadrature $I$. From a double gaussian fit to the marginal distribution along the $I$ quadrature, we extract the means $\mu_\text{g,e}\approx \pm 3.0\,\sqrt{\text{photon}}$ and standard deviations $\sigma_\text{g,e} \approx 1.0 \,\sqrt{\text{photon}}$ for the qubit states. To assign the qubit state to the demodulated contiguous data, we use a two-point latching filter, which declares a change in the qubit state when the in-phase value $I$ enters the $\mu \pm 2\sigma$ band (\figrefadd{fig:supp:quantumjumps}{a}) centered on the other qubit state. The extracted population of the excited state corresponds to an effective qubit temperature of \SI{37}{\milli\kelvin}. Moreover, by histogramming the lifetime of the states (\figrefadd{fig:supp:quantumjumps}{b,c}), we find the energy relaxation during readout $T_1=(T_\downarrow^{-1}+T_\uparrow^{-1})^{-1}=\SI{9.8}{\micro\second}$, practically unchanged from the free decay value (\figrefadd{fig:timedomain}{a}). 

\section{Evidence of \texorpdfstring{$E_\text{J}$}{EJ} Toggling in the Gralmonium Spectrum vs. Flux}
\label{sec:supp:splitspectrum}
\begin{figure*}[t]
\centering
\includegraphics[width=\textwidth]{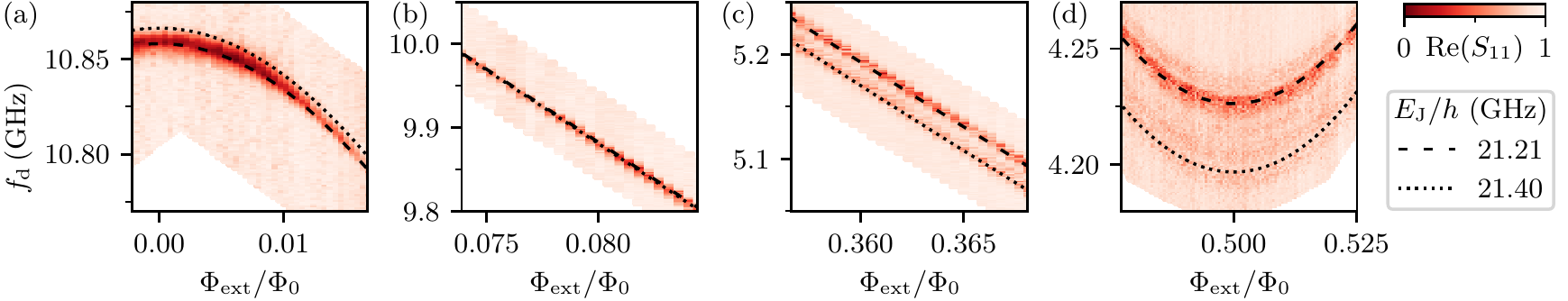}
\caption{\textbf{Evidence of $E_\text{J}$ toggling in the gralmonium spectrum vs. flux.}
The panels show spectroscopy of the same gralmonium as in \figref{fig:spectrum}, but in a previous cooldown ($\#3$) without the use of a parametric amplifier. Each trace is averaged over \revisex{$\SI{30}{\second}$} \revise{\SI{50}{\second}} and contains 100 points. The color scale corresponds to the in-phase component of the single-port reflection coefficient $S_{11}$, with the minimum and maximum values rescaled to the $0-1$ interval for clarity.  Two dominant qubit frequencies are visible across the entire flux range and can be fitted (dashed and dotted line) with identical circuit parameters except for the values of the nano-junction $E_\text{J}$, which differ by \SI{190}{\mega\hertz}. Close to zero flux (\textbf{a}) the difference between the two spectra is \SI{7.4}{\mega\hertz}. As expected from the model, the lines cross at $\Phi_\text{ext}/\Phi_0\approx 0.08$ (\textbf{b}) and they are again visibly separated towards half-flux (\textbf{c}). The largest splitting (\SI{30}{\mega\hertz}) occurs at half-flux (\textbf{d}), where additional lines also become visible.
}
\label{fig:supp:split}
\end{figure*}
As discussed in the main text and in \figref{fig:timedomain}, the gralmonium frequency toggles even when kept at cryogenic temperatures. This toggling is also visible in continuous wave spectroscopy versus flux when the trace averaging time is \revise{comparable to the respective timescale of the toggling.}\revisex{ longer than minutes;} \revise{In the half flux spectroscopy discussed in the main text (\figrefadd{fig:spectrum}{b} inset), the toggling on minutes timescale is visible as jumps in the qubit frequency every few traces. In \figref{fig:supp:split}, we show spectroscopy in a previous cooldown without a parametric amplifier, where the averaging time per trace is about one order of magnitude longer. Therefore, the toggling on minutes timescale is imprinted on the spectroscopy data as distinct qubit lines associated to the qubit transitions. In contrast to the jumps between traces in \figrefadd{fig:spectrum}{b}, the distinct qubit lines are visible in \figref{fig:supp:split} within the same trace.} The observation of two qubit frequencies across the entire flux range rules out a transverse coupling to parasitic two level systems (TLSs) residing at fixed frequency. Indeed, the two qubit lines are captured by two numerical fits to the fluxonium Hamiltonian, which differ only in the value of the Josephson energy $E_\text{J}$. Notably, the two-$E_\text{J}$-model correctly predicts the merging of the two frequencies at $\Phi_\text{ext}/\Phi_0\approx 0.08$ (\figrefadd{fig:supp:split}{b}), as well as their frequency inversion between zero and half-flux biases. 

\revise{
\section{Power Spectral Density}
\label{sec:supp:PSD}
\begin{figure}[t]
\centering
\includegraphics[width=\columnwidth]{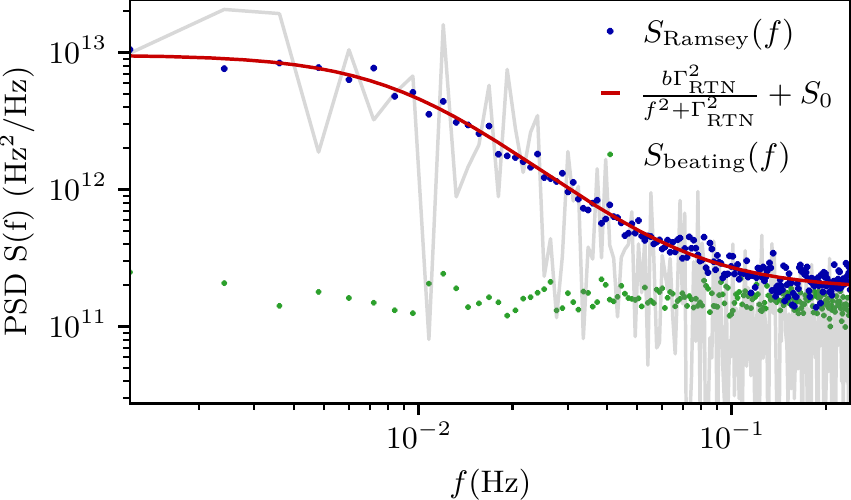}
\caption{\revise{
\textbf{Power spectral density $S(f)$ of the gralmonium frequency on the minutes timescale.}
The power spectral density is calculated from a timetrace of $400$ measurements of the gralmonium frequency extracted from contiguous Ramsey experiments, with a total duration of \SI{832}{\second} (example in light gray). This sequence is repeated 85 times and the blue markers show the averaged  power spectrum. The power spectrum $S_\text{Ramsey}(f)$ follows a Lorentzian shape at low frequencies, corresponding to random telegraphic noise (RTN). A fit (red line) reveals the switching rate $\Gamma_\text{RTN} = \SI{9.4}{\milli\hertz}$, amplitude $b=\SI{1.89e13}{\square\hertz /\hertz}$ and white noise floor of $S_0 =\SI{3.73e11}{\square\hertz /\hertz}$. For comparison, the power spectrum of the beating frequency $f_\text{beating}$ (cf. \figrefadd{fig:timedomain}{d}) obtained from the same measurements only consists of a white noise component (cf. green markers).
}}
\label{fig:supp:psd}
\end{figure}
For a quantitative analysis of the toggling of the qubit frequency on minutes timescale, we calculate the power spectral density 
\begin{align}
    S(f) = \mathcal{F}^2(f_\text{01}) \cdot 2T\,,
\end{align}
where $\mathcal{F}(f_\text{01})$ is the normalized discrete Fourier transform and $T$ is the total measurement duration. From double-frequency fits (cf.~\figrefadd{fig:timedomain}{d}) to $400$ contiguous Ramsey measurements, we extract a time series of the Ramsey frequency $f_\text{Ramsey}$ and the beating frequency $f_\text{beating}$ with a total duration $T=\SI{832}{\second}$. \Figref{fig:supp:psd} shows the power spectrum for $f_\text{Ramsey}$ (blue markers) and $f_\text{beating}$ (green markers) averaged over 85 traces. The power spectrum of $f_\text{Ramsey}$ follows a Lorentzian shape characteristic for random telegraphic noise (RTN),
\begin{align}
    S(f) = \frac{ b\Gamma^2 _\text{RTN}}{f^2 + \Gamma^2 _\text{RTN}}+S_0 \label{eq:supp:PSDfit}\,,
\end{align}
where $\Gamma_\text{RTN}$ is the switching rate, $b$ the amplitude and $S_0$ the frequency independent white noise floor. A fit to $S_\text{Ramsey}(f)$ (red line) reveals a switching rate of $\Gamma_\text{RTN} = \SI{9.4}{\milli\hertz}$. In contrast, the power spectrum of the beating frequency $f_\text{beating}$ is frequency independent.
}

\section{Additional Gralmonium Spectra}
\label{sec:supp:othersamples}
\label{sec:supp:paramtersvscd}
\begin{figure*}[p]
\centering
\includegraphics[width=\textwidth]{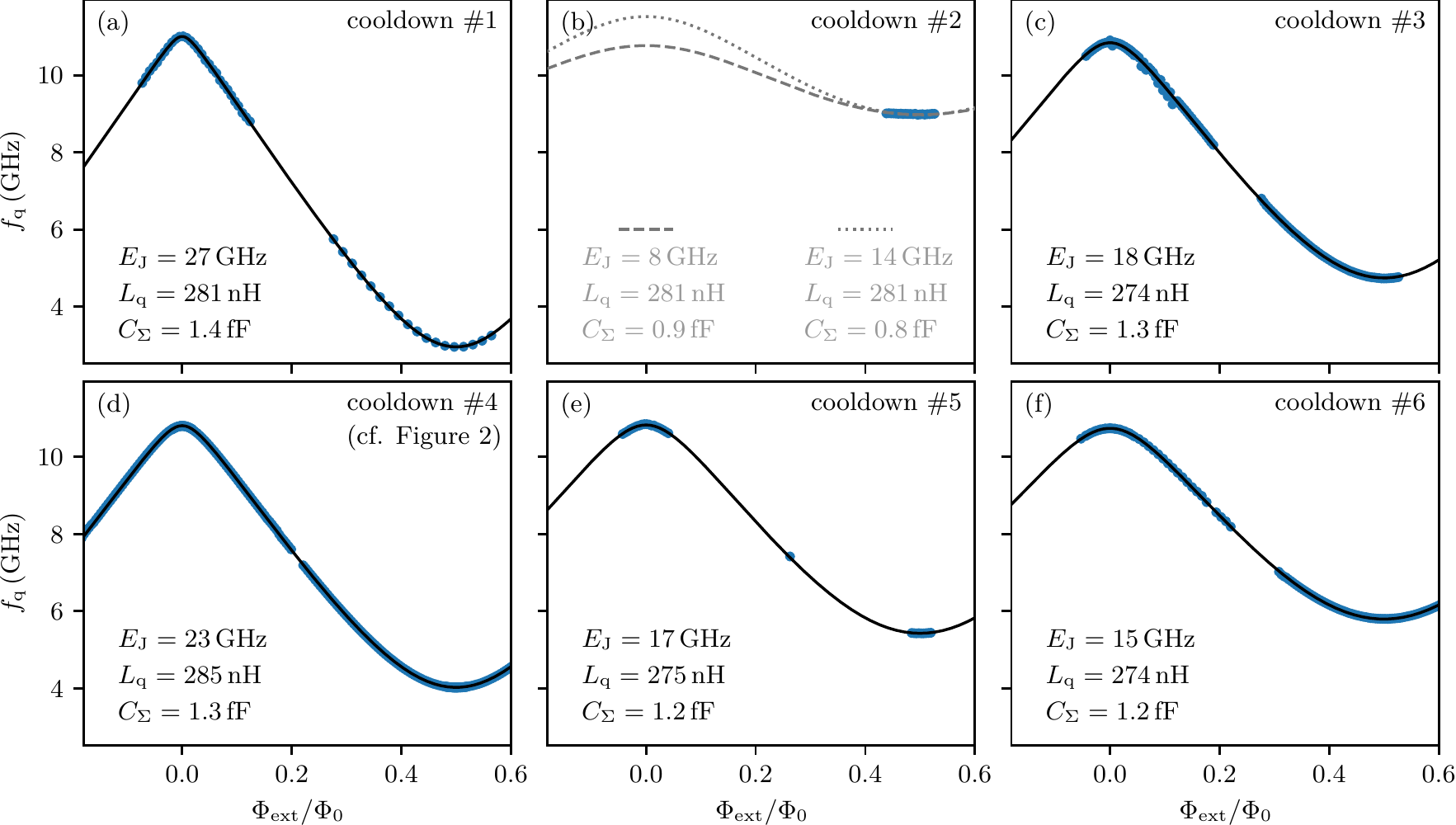}
\caption{
\textbf{Spectra and fit parameters in consecutive cooldowns of the main sample.}
The blue markers are extracted from spectroscopy of the $\ket{g}\to\ket{e}$ transition and the black lines show numerical fits of \eqref{eq:fluxoniumHamiltonian} to the data. The spectrum undergoes significant changes between cooldowns, particularly visible at half-flux. The corresponding changes in half-flux frequency and $E_\text{J}$ are summarized in \figrefadd{fig:timedomain}{e}. A comparison of the fit parameters reveals relatively large changes in $E_\text{J}$ and moderate changes in $C_\Sigma$. In contrast, the inductance $L_\text{q}$ remains constant within few percent. For the second cooldown (panel \textbf{(b)}) only data around half-flux was taken, making an unambiguous fit difficult. Instead, we show two plausible parameter sets by fixing $L_\text{q}$ to the value of the previous cooldown and $C_\Sigma$ around the lower bound $C_\text{q}\gtrsim\SI{0.8}{\femto\farad}$ expected from the finger capacitance and observed in other samples (cf. \figrefadd{fig:supp:othersamples}{a}).
}
\label{fig:supp:otherCD}
\end{figure*}
The gralmonium sample discussed in the main text was measured in 6 consecutive cooldowns (\figref{fig:supp:otherCD}), revealing significantly different spectra. From fits (black lines) to the fluxonium Hamiltonian (\eqref{eq:fluxoniumHamiltonian}) we identify the nano-junction $E_\text{J}$ as the parameter which changes the most between cooldowns. The total gralmonium capacitance $C_\Sigma = C_\text{q} + C_\text{J}$ also changes and appears to be correlated with $E_\text{J}$, however, it is bounded by the finger capacitance $C_\text{q}\approx \SI{0.8}{\femto\farad}$. Therefore, the gralmonium charging energy is limited to values below $E_\text{C}^\Sigma \leq E_\text{C}^\text{q} \approx \SI{24}{\giga\hertz}$.

\begin{figure*}[p]
\centering
\includegraphics[width=\textwidth]{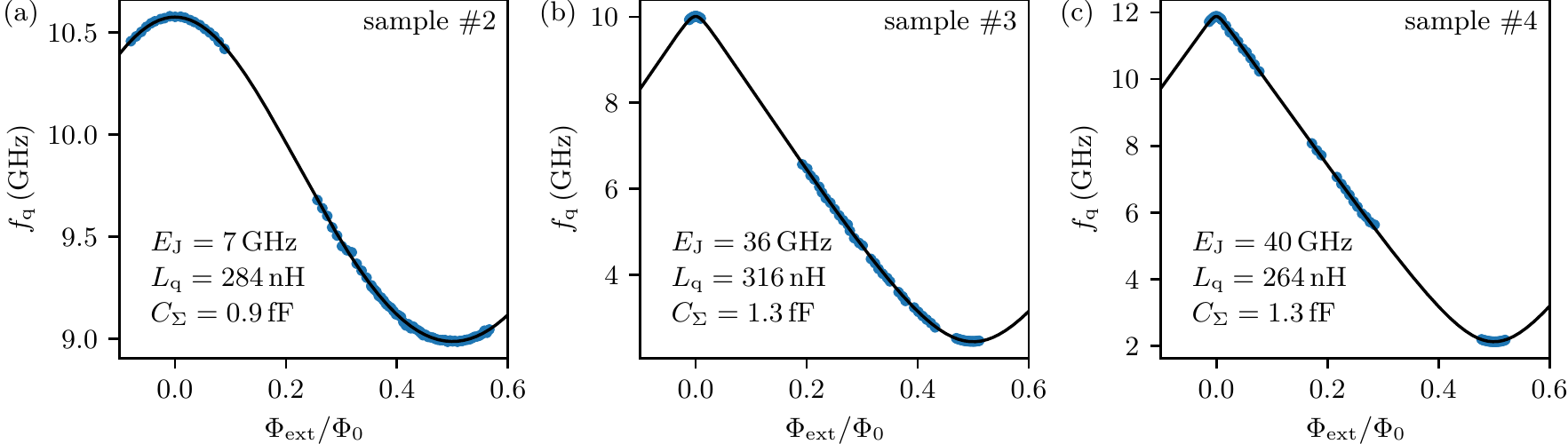}
\caption{
\textbf{Spectra and parameters of additional gralmonium samples}. The blue markers are extracted from spectroscopy of the $\ket{g}\to\ket{e}$ transition and the black lines show numerical fits of \eqref{eq:fluxoniumHamiltonian} to the data. \textbf{(a)} The Josephson energy of some grAl nano-junctions is small enough to lift the entire spectrum of the first transition to the \SI{10}{\giga\hertz} range. Samples \textbf{(b)} and \textbf{(c)} feature similar parameters as the gralmonium in \figref{fig:spectrum}.
}
\label{fig:supp:othersamples}
\end{figure*}
\revise{Including the sample characterized in the main text, we measured the spectra of 20 gralmonium devices across 11 wafers} and all data is consistent with the standard fluxonium Hamiltonian \eqref{eq:fluxoniumHamiltonian}. In \figref{fig:supp:othersamples} we show the spectra and fit parameters of three selected devices with the same circuit design as in \figref{fig:sample}. The spread of the Josephson energy across devices\revise{, \SIrange{7}{40}{\giga\hertz},} is similar to variations of individual nano-junctions in successive cooldowns (\revise{\SIrange{8}{27}{\giga\hertz},} cf.~\figref{fig:supp:otherCD}).

\end{document}